\documentclass[twocolumn,showpacs,amsmath,amssymb,10pt]{revtex4}
\usepackage{graphicx,color}
\usepackage{amsmath}
\usepackage{amssymb}
\usepackage{bbm}
\usepackage{amsfonts}
\usepackage{bm}

\newcommand{\ot}{{\,\otimes\,}}
\newcommand{{\Cd}}{{\mathbb{C}^d}}

\def\oper{{\mathchoice{\rm 1\mskip-4mu l}{\rm 1\mskip-4mu l}
{\rm 1\mskip-4.5mu l}{\rm 1\mskip-5mu l}}}
\def\<{\langle}
\def\>{\rangle}

\newcommand{\beq}{\begin{equation}}
\newcommand{\eeq}{\end{equation}}
\newcommand{\bear}{\begin{eqnarray}}
\newcommand{\ear}{\end{eqnarray}}
\newcommand{\bdm}{\begin{displaymath}}
\newcommand{\edm}{\end{displaymath}}

\begin{document}
\title{\textbf{
Feshbach projection formalism for open quantum systems
}}
\author{Dariusz Chru\'sci\'nski and Andrzej Kossakowski }
\affiliation{ Institute of Physics \\ 
Nicolaus Copernicus University \\
Grudzi{a}dzka 5, 87--100 Torun, Poland\\
}

\begin{abstract}
We provide a new approach to open quantum systems which is based on the Feshbach projection method. Instead of looking for a master equation for the dynamical map acting in the space of density operators we provide the corresponding equation for the evolution in the Hilbert space of the amplitude operators. Its solution enables one to construct a legitimate quantum evolution (completely positive and trace preserving). Our approach, contrary to the standard Nakajima-Zwanzig method, allows for a series of consistent approximations resulting in a legitimate quantum evolution. The new scheme is illustrated by the well known spin-boson model beyond rotating wave approximation. It is shown that the presence of counter-rotating terms dramatically changes the asymptotic evolution of the system.


\end{abstract}

\pacs{03.65.Yz, 03.65.Ta, 42.50.Lc}

\maketitle


{\em Introduction. --} The description of a quantum system interacting with its environment is of fundamental importance for quantum physics and defines the central objective of the theory of open quantum systems \cite{Breuer,Weiss}. During the last few years there has been an increasing interest in  open quantum systems in connection to the growing interest in controlling quantum systems and applications in modern quantum technologies such as quantum communication, cryptography,  computation and ever growing number of applications.
In practice, this theory is usually applied in the so-called Markovian or memoryless approximation. However, when
strong coupling or long environmental relaxation times make memory effects
important for a realistic description of the dynamics one needs more refined approach and hence the general structure of non-Markovian quantum evolution is a crucial issue \cite{Wolf,RHP,BLP,PRL}. For the recent papers devoted to both theoretical and experimental aspects of quantum evolution with memory see e.g. a collection of papers in  \cite{REV} and references therein.



The standard approach to the dynamics of open system uses the
Nakajima-Zwanzig projection operator technique \cite{NZ} which shows
that under fairly general conditions, the master equation for the
reduced density matrix takes the form of the following
non-local equation
\begin{equation}\label{NZ}
\frac{d}{dt}\rho_t = \int_{0}^t \mathcal{K}_{t-u}\rho_u\,
du\ ,
\end{equation}
in which quantum memory effects are taken into account through the
introduction of the memory kernel $\mathcal{K}_t$: this simply
means that the rate of change of the state $\rho_t$ at time $t$
depends on its history.  An
alternative and technically much simpler scheme is provided by the
time-convolutionless 
projection operator technique
\cite{TCL,BKP,Breuer} in which one obtains a first-order
differential equation for the reduced density matrix
\begin{equation}\label{local}
\frac{d}{dt}\rho_t = L_t \rho_t\ .
\end{equation}
 The advantage
of the local approach consists in the fact that it yields an equation
of motion for the relevant degrees of freedom which is local in time
and which is therefore often much easier to deal with than the
Nakajima-Zwanzig non-local master equation (\ref{NZ}).

It should be stressed that the structure of the memory kernel $\mathcal{K}_t$ is highly nontrivial and, therefore, the non-local master equation (\ref{NZ}) is rather untractable. Note, that this equation is exact, i.e. in deriving (\ref{NZ}) one does not use any specific approximation. Approximating (\ref{NZ}) is a delicate issue. One often applies second order Born approximation which considerably simplifies  the structure of $\mathcal{K}_t$. However, this approximation in general violates basic properties of the master equation like for example complete positivity or even positivity of $\rho_t$ \cite{Fabio-rev}. Further simplification of (\ref{NZ}) consists in various Markov approximations which allow one to avoid memory effects. These approximations may also break the physics of the problem. For example well known local Redfield equation \cite{Redfield} again violates complete positivity \cite{Breuer,Fabio-rev}. The problem of a consistent Markov approximation was studied in \cite{Dumcke}. One often tries to use phenomenological memory kernels. However, as was already observed in \cite{Stenholm}, there is no simple recipe how to construct $\mathcal{K}_t$ in order to preserve basic properties of quantum evolution \cite{Lidar,Sabrina}.

The local approach based on  (\ref{local})  is much more popular
and provides a straightforward generalization of the celebrated Markovian semigroup \cite{GKS,Lindblad}.
%
%
It should be also stressed that Markovian semigroup, being a special case of (\ref{local}), is derived from (\ref{NZ}) by applying quite sophisticated Markovian approximations like for example weak coupling or singular coupling limits \cite{Dumcke,Alicki}.

In this Letter we provide a new approach to the reduced dynamics of open quantum systems. Instead of applying the Nakajima-Zwanzig projection we apply the Feshbach projection formalism \cite{Feshbach} to the Schr\"odinger equation of the total system. This formalism was recently applied in the context of open quantum system in \cite{Gaspard,Wu,Wu1}. In this Letter we use Feshbach projection technique to derive a closed formula for the reduced dynamics (see formula (\ref{Lambda-S})). We stress that although the Feshbach projection technique is well known the above formula for the dynamical map $\Lambda_t$ is completely new. We illustrate the power of this method analyzing spin-boson model beyond rotating wave approximation (RWA). The big advantage of this approach is the ability of performing a consistent approximation which creates notorious  problems in the standard Nakajima-Zwanzig approach. However, the essential limitation of this method is that the initial state of the environment has to be pure (for example in the standard spin-boson model one starts with the vacuum state of the boson field \cite{Breuer}). To get rid of this constraint we propose a {\em generalized Feshbach projection method} which enables one to start with an arbitrary mixed state of the environment. This generalized technique allows to analyze spin-boson model beyond RWA and with arbitrary mixed state of the field. As a byproduct we derive a new description of quantum systems based not on the density matrix $\rho$ but on the amplitude operator
$\kappa$ satisfying $\rho =\kappa \kappa^\dagger$. Clearly, $\kappa$ is not uniquely defined (it is gauge dependent) but the whole theory is perfectly gauge invariant.



{\em Feshbach projection technique. --} Consider  a quantum system coupled to the environment living in $\mathcal{H}_S\ot \mathcal{H}_E$ and  let $H$ denote the total Hamiltonian of the composed system
\begin{equation}\label{}
    H = H_0 + V = H_S \ot \mathbb{I}_E + \mathbb{I}_S \ot H_E + V \ .
\end{equation}
Passing to the interaction picture $V(t) = e^{iH_0 t} V e^{-i H_0 t}$ one considers
\begin{equation}\label{H-Psi}
    i \partial_t \Psi_t = V(t) \Psi_t\ .
\end{equation}
Now, let $\psi_E$ be a fixed vector state of the environment and let us introduce an orthogonal projector $P_0 : \mathcal{H}_S\ot \mathcal{H}_E \rightarrow \mathcal{H}_S\ot \mathcal{H}_E$ defined by
$$P_0 \psi \ot \phi = \psi \ot \psi_E \< \psi_E|\phi\>\ , \ \ \  $$
and by linearity one defines $P_0 \Psi$ for arbitrary vector $\Psi$. Moreover, let $P_1 = \mathbb{I}_S \ot \mathbb{I}_E - P_0$ denotes a complementary projector. The standard projection technique gives
\begin{eqnarray} \label{Rel}
 \partial_t P_0 \Psi_t  &=& -i V_{00}(t) P_0 \Psi_t - i V_{01}(t) P_1 \Psi_t\ , \\  \label{Irr}
  \partial_t P_1 \Psi_t &=& -i V_{10}(t) P_0 \Psi_t - i V_{11}(t) P_1 \Psi_t\ ,
\end{eqnarray}
where we introduced a convenient notation $V_{ij}(t) = P_i V(t) P_j$. Assuming separable initial state $\Psi_0 = \psi \ot \psi_E$ and solving  (\ref{Irr}) for the irrelevant part $P_1\Psi_t$
\begin{equation}  \label{P1}
    P_1\Psi_t = - i \int_0^t ds\, W_{t,s} V_{10}(s) P_0 \Psi_s\ ,
\end{equation}
one ends up with the following non-local equation for the relevant (system) part $P_0\Psi_t$:
\begin{equation}\label{R-P}
     \partial_t P_0\Psi_t = -i V_{00}(t) P_0\Psi_t - \int_0^t {K}_{t,s} P_0 \Psi_s\, ds\ ,
\end{equation}
with ${K}_{t,s} = V_{01}(t) W_{t,s} V_{10}(s)$, and
\begin{equation}\label{W-ts}
    W_{t,s} = \mathcal{T}\, \exp\left( -i \int_s^t V_{11}(u) du \right)\ ,
\end{equation}
where $\mathcal{T}$ denotes chronological product. Let $Z_t : \mathcal{H}_S \rightarrow \mathcal{H}_S$ be defined by
\begin{equation}\label{}
    (Z_t \psi) \ot \psi_E = P_0 \Psi_t = P_0 U_t (\psi \ot \psi_E) \ ,
\end{equation}
where  $U_t$ provides a solution to the original Schr\"odinger equation (\ref{H-Psi}), that is, $i\partial_t U_t = V(t)U_t$. Equation (\ref{R-P}) may be rewritten as the following equation for $Z_t$
\begin{equation}\label{R}
     \partial_t Z_t = -i V_{\rm eff}(t) Z_t - \int_0^t M_{t,s} Z_s\, ds\ ,
\end{equation}
where the effective  time-dependent system Hamiltonian is defined by $V_{\rm eff}(t) = {\rm tr}_E(V(t) \, \mathbb{I}_S\ot |\psi_E\>\<\psi_E|)$ and $M_{t,s} = {\rm tr}_E(\mathbb{K}_{t,s} \, \mathbb{I}_S \ot |\psi_E\>\<\psi_E|)$. Solving (\ref{R}) one finds the reduced evolution of initial state vector $\psi \in \mathcal{H}_S$: $\psi \rightarrow \psi_t = Z_t \psi_0$. Let us observe that $\<Z_t\psi|Z_t\psi\> \leq \<\psi|\psi\>$ which shows that $Z_t \psi$ is no longer a legitimate vector state for $t > 0$. It is clear since $Z_t$ describes the decay of $\psi$ and hence $||Z_t \psi||$ is not conserved -- it leaks out to the irrelevant part $P_1U_t(\psi\ot \psi_E)$.
On the other hand there is a standard formula for the dynamical map
\begin{equation}\label{Lambda}
\Lambda_t(|\psi\>\<\psi|) = {\rm tr}_E [ U_t (|\psi\>\<\psi| \ot |\psi_E\>\<\psi_E|) U_t^\dagger]\ .
\end{equation}
Simple calculation  shows that inserting the identity $\mathbb{I}_S \ot \mathbb{I}_E=P_0 + P_1$ under the partial trace  $[(P_0+P_1) U_t (|\psi\>\<\psi| \ot |\psi_E\>\<\psi_E|) U_t^\dagger(P_0+P_1)]$ and using the definition of $Z_t$ and formula (\ref{P1})  one obtains
\begin{equation}\label{Lambda-S}
    \Lambda_t(|\psi\>\<\psi|) = Z_t \rho Z_t^\dagger + {\rm Tr}_E ( Y_t [\rho \ot |\psi_E\>\<\psi_E|] Y_t^\dagger)\ ,
\end{equation}
where the operator $Y_t$ 
is defined by
\begin{equation}\label{}
    Y_t = \int_0^t ds \ W_{t,s} V_{10}(s) (Z_s \ot \mathbb{I}_E)  \ .
\end{equation}
By linearity one defines the action of $\Lambda_t$ on an arbitrary density operator: if $\rho = \sum_k p_k |\psi_k\>\<\psi_k|$, then
$\Lambda_t(\rho) = \sum_k p_k \Lambda_t(|\psi_k\>\<\psi_k|)$.
It should be stressed that although the Feshbach projection technique is well known  \cite{Gaspard,Wu,Wu1} the above formula for the reduced dynamics is completely new. Note, that presented method requires that the initial state of the environment is a pure  vector state $\psi_E$. Hence the standard Feshbach projection method is much more restrictive  than the corresponding Zwanzig-Nakajima method. However, the advantage of the Feshbach technique consists in the fact that (contrary to the Zwanzig-Nakajima method)  it allows for consistent approximations. By a consistent we mean an approximation which results in completely positive and trace preserving evolution of a density matrix.


{\em Born-like approximation. --} Note, that the original memory kernel $M_{t,s}$ contain an infinite number of multi-time correlations functions which makes the full problem rather untractable.
The simplest approximation consists in neglecting $V_{11}(t)$.
Roughly speaking $V_{11}(t)$ is responsible for transitions within irrelevant part of the Hilbert space $P_1(\mathcal{H}_S \ot \mathcal{H}_E)$. It means that one approximates the evolution operator $W_{t,s}$ by $\mathbb{I}_S \ot \mathbb{I}_E$.
It leads to the second order approximation for the memory kernel ${M}_{t,s} \simeq {\rm Tr}_E [\, \mathbb{I}_S\ot |\psi_E\>\<\psi_E|  V_{01}(t)V_{10}(s)\, ]$ and hence it is natural to call it Born-like approximation.
The big advantage of our approach consists in the fact that the above approximation leads to the legitimate completely positive and trace preserving quantum evolution.



{\em Example: spin-boson model. --} To illustrate our approach let us consider well known spin-boson model \cite{Breuer} beyond RWA   defined by
\begin{equation}\label{}
    H_S = \omega_0 \sigma^+\sigma^- \ , \ \ \ H_E = \int dk\, \omega(k) a^\dagger(k) a(k)\ ,
\end{equation}
and the interaction term
\begin{equation}\label{}
    V = \sigma^+ \ot X + \sigma^- \ot X^\dagger\ ,
\end{equation}
where $X = a(f) + a^\dagger(h) $,  and $a^\dagger(f) = \int dk\, f(k) a^\dagger(k)$. As usual $\sigma^\pm$ are standard raising and lowering qubit operators. Note that a form-factor `$h$' introduces counter-rotating terms. One easily computes
\begin{equation}\label{}
    V(t) = \sigma^+ \ot X(t) + \sigma^- \ot X^\dagger(t)\ ,
\end{equation}
where $X(t) = e^{-i \omega_0 t} [a(f_t) + a^\dagger(h_t)]$, and
the time-dependent form-factors read $f_t(k) = e^{-i \omega(k)t} f(k)$  and a similar formula for $h_t$.  Recall that in the standard spin-boson model $h=0$ (no counter-rotating terms) and $\psi_E = |{\rm vac}\>$ is the vacuum state of the boson field \cite{Breuer}.
Let $\mathcal{H}_0$ be a 2-dimensional subspace of $\mathcal{H}_S \ot \mathcal{H}_E$ spanned by $|i\> \ot |{\rm vac}\>$ for $i =1,2$, and $\mathcal{H}_1$ a subspace spanned by $|1\> \ot a^\dagger(k)|{\rm vac}\>$. One takes the initial state $\Psi_0 = |\psi\> \ot |{\rm vac}\> \in \mathcal{H}_0$. The structure of the interaction Hamiltonian $V(t)$ within RWA guaranties that $\Psi_t \in \mathcal{H}_0 \oplus \mathcal{H}_1$. One easily finds
 $   P_1 V(t) P_1 \Big|_{\mathcal{H}_0 \oplus \mathcal{H}_1} = 0$,
which shows that the Born-like approximation within the standard spin-boson model is exact.
Actually, due to this fact the standard model is exactly solvable.

Consider now the spin-boson model beyond RWA and let $\psi_E$ be a fixed pure state of the environment. If $B_1,\ldots,B_n$ are field operators, then denote by $\< B_1\ldots B_n\> = \<\psi_E|B_1 \ldots B_n|\psi_E\>$ the corresponding correlation function. To simplify our presentation let us assume that $\psi_E$ satisfies
 $   \< a(f) \> = \< a^\dagger(f)\> = 0$.
The above condition is satisfied for the vacuum state. It is clear that this condition implies $\< X(t)\>=0$, and hence $V_{\rm eff}(t)=0$. In the Born-like approximation the formula for ${M}_{t,s}$ reduces to
\begin{eqnarray}\label{}
    {M}_{t,s} 
    = m_1(t,s)\, |1\>\<1|  + m_2(t,s)\, |2\>\<2|\ ,
\end{eqnarray}
with
$m_1(t,s) = \< X(t) X^\dagger(s) \>$ and $m_2(t,s) =  \< X^\dagger(t) X(s) \>$.
which proves that within this approximation the dynamics of ${Z}_t$ is fully controlled by 2-point correlation functions.
 It is, therefore, clear that $Z_t$ has the following form
\begin{equation}\label{}
{Z}_t = {z}_1(t) |1\>\<1| + {z}_2(t) |2\>\<2|\ ,
\end{equation}
where the complex functions ${z}_k(t)$ satisfy
\begin{eqnarray}
  \partial_t {z}_k(t)  = - \int_0^t m_k(t,s)\, \, {z}_k(s)\, ds\ ,
\end{eqnarray}
with ${z}_k(0)=1$. Interestingly, we have two decoupled equations for $z_k$. Having solved for ${Z}_t$ one computes a second part of the dynamical map (\ref{Lambda-S}), namely ${\rm Tr}_E ( {Y}_t [\,\rho \ot |\psi_E\>\<\psi_E|] {Y}_t^\dagger)\,$, where in the Born-like approximation ${Y}_t = \int_0^t ds {V}_{10}(s) {Z}_s \ot \mathbb{I}_E$.
Observing that ${V}_{10}(s) = {P}_1 {V}(s) {P}_0 = {V}(s){P}_0$ due to ${P}_0V(t){P}_0=0$, one finds the following Kraus representation for the dynamical map
\begin{eqnarray*}
\Lambda_t(\rho) &=& {Z}_t \rho {Z}_t^\dagger +  d_1(t) \sigma^+\rho \sigma^- + d_2(t) \sigma^-\rho \sigma^+  \\
&+& \alpha(t) \sigma^+\rho \sigma^+ + \alpha^*(t) \sigma^-\rho \sigma^-\ ,
\end{eqnarray*}
where
\begin{eqnarray*}
  d_1(t) &=& \int_0^t ds \int_0^t du \, \< X^\dagger(u) X(s) \> \, z_1(s) z_1^*(u)   \ , \\
  d_2(t) &=& \int_0^t ds \int_0^t du \, \< X(u) X^\dagger(s) \>\,  z_2(s) z_2^*(u)  \ , \\
  \alpha(t) &=& \int_0^t ds \int_0^t du \, \< X(u) X(s) \>\, z_1(s) z_2^*(u) \ .
\end{eqnarray*}
Interestingly, the preservation of trace implies that $ d_k(t) + |z_k(t)|^2 = 1$,
for $k=1,2$. Actually, this simple condition is hardly visible from the definition of $d_k(t)$ and the corresponding non-local equations for $z_k(t)$. The evolution of density matrix $\rho_{ij}(t)$ reads:
\begin{eqnarray}
  \rho_{11}(t) &=& |z_1(t)|^2 \rho_{11} + d_2(t) \rho_{22} \ ,\nonumber  \\
  \rho_{22}(t) &=& d_1(t) \rho_{11} + |z_2(t)|^2 \rho_{22}\ ,  \\
  \rho_{12}(t) &=& z_1(t)z_2^*(t) \rho_{12} + \alpha(t) \rho_{21} \ . \nonumber
\end{eqnarray}
Interestingly, if $|z_1(t)|^2$ and $|z_2(t)|^2$ vanish at infinity then asymptotically one has
\begin{equation}\label{}
    \rho_{11}(t) \rightarrow \rho_{22} \ , \ \ \   \rho_{22}(t) \rightarrow \rho_{11}  \ ,
\end{equation}
which means that the occupations $\rho_{11}$ and $\rho_{22}$ simply swap. Hence, such evolution does not have a proper equilibrium state (its asymptotic state highly depend upon the initial one). Such asymptotic behavior is excluded in the standard spin-boson model without counter-rotating terms corresponding to $|z_1(t)|=1$ and $\alpha(t)=0$. Again, if $|z_2(t)|^2$ vanish at infinity then asymptotically $ \rho_{11}(t) \rightarrow 1$ and  $\rho_{22}(t) \rightarrow 0$ which means that the ground state $|1\>\<1|$ defines the unique equilibrium  state. Our analysis shows that the presence of anti-resonant terms dramatically changes the asymptotic evolution of the system since the system does not possesses an equilibrium state.


{\em Density operators vs. amplitudes. --} It should be stressed that presented method requires that the initial state of the environment has to be pure. There is no natural way within presented approach to generalize formula (\ref{Lambda-S}) for arbitrary mixed state of the environment. To get rid of this limitation we provide a new approach to the dynamics of  quantum systems.
Our approach is based not on the Schr\"odinger equation for the vector state $|\Psi_t\>$ but on the Schr\"odinger-like equation for the ``amplitude'' operator. Let as recall that if $\rho$ is density operator in $\mathcal{H}$ then a Hilbert-Schmidt operator $\kappa \in \mathcal{L}^2(\mathcal{H})$ is called an amplitude of $\rho$ if $\rho =  \kappa\kappa^\dagger$. Recall, that $\mathcal{L}^2(\mathcal{H})$ is equipped with a scalar product $(\kappa,\eta) = {\rm tr}(\kappa^\dagger \eta)$ and $\kappa \in \mathcal{L}^2(\mathcal{H})$ if the Hilbert-Schmidt norm $||\kappa||^2 = (\kappa,\kappa)$ is finite. Note that ${\rm tr}\, \rho=1$ implies $(\kappa,\kappa)=1$. If $a^\dagger = a$ is an observable, then
\begin{equation}\label{}
    (\kappa,a\, \kappa) = {\rm tr}\, (\kappa^\dagger a \kappa) = {\rm tr}\, ( a \kappa\kappa^\dagger)={\rm tr}\,(a \rho)\ ,
\end{equation}
reproduces the standard formula for the expectation value of $a$ in the state $\rho$. Amplitudes display a natural gauge symmetry: a gauge transformation $\kappa \rightarrow \kappa \, U$ leaves $\rho=\kappa \kappa^\dagger$ invariant for any unitary operator $U$ in $\mathcal{H}$.
The main idea of this paper is to analyze the dynamics of gauge invariant $\rho$ in terms of its gauge dependent amplitudes $\kappa$. This is well known trick in physics. Recall that for example in Maxwell theory it is much easier to analyze Maxwell equations not in terms of gauge invariant $\mathbf{E}$ and $\mathbf{B}$ fields but in terms of gauge dependent four potential $A_\mu$. Suppose that $\rho_t$ satisfies von Neumann equation
 $   i\partial_t {\rho}_t = [H_t,\rho_t]$,
with time-dependent Hamiltonian $H_t$ (it might be for example the interaction Hamiltonian in the interaction picture). Its solution is given by $\rho_t = v_t \rho v_t^\dagger$, where the unitary operator $v_t$ solves the Schr\"odinger equation $i \partial_t v_t = H_t v_t$ with the initial condition $v_0 = \mathbb{I}$. The corresponding equation for the amplitude $\kappa_t$ is highly non unique.
Any equation
  $  i\partial_t {\kappa}_t = H_t\kappa_t - \kappa_t G_t$,
where $G_t$ is an arbitrary time-dependent Hermitian operator, does the job. The quantity $G_t$ plays a role of a gauge field. It is clear that the solution $\kappa_t$ does depend upon $G_t$ but $\rho_t = \kappa_t \kappa_t^\dagger$ is perfectly gauge-invariant. If $G_t=H_t$ then $\kappa_t$ satisfies the same von-Neumann equation as $\rho_t$ (such choice is used e.g. in \cite{Froehlich}).
Taking $G_t=0$ one arrives at the following  Schr\"odinger-like equation for the amplitude.
\begin{equation}\label{vN-kappa-}
    i\partial_t {\kappa}_t = H_t\kappa_t \ .
\end{equation}
As we shall see this simple choice  leads to considerable simplification of the underlying structure of the gauge theory. Note that
the Schr\"odinger-like equation (\ref{vN-kappa-}) still allows for global (i.e. time independent) gauge transformations $\kappa_t \rightarrow \kappa_t u$. Equation (\ref{vN-kappa-}) provides a starting point for the generalized Feshbach method. We show that one may replace in (\ref{Lambda-S}) a pure state $|\psi_E\>\<\psi_E|$ by an arbitrary mixed state of the environment. To justify this statement we shall work with amplitudes instead of density operators.

{\em Generalized Feshbach projection technique. --} We generalize the Feshbach projection method to the Hilbert space of amplitude operators $\mathcal{L}^2(\mathcal{H}_S \ot \mathcal{H}_E)$. Let $\Omega$ be a fixed state of the environment and let $\kappa_E$ be its amplitude. Let us introduce orthogonal  projectors
\begin{equation}\label{}
    \mathcal{P}_0 \mu_S\ot \mu_E = \mu_S \ot \kappa_E(\kappa_E,\mu_E)\ ,\ \ \ \mathcal{P}_1 = \oper - \mathcal{P}_0\ .
\end{equation}
where $\mu_S \in \mathcal{L}^2(\mathcal{H}_S)$ and $\mu_E \in \mathcal{L}^2(\mathcal{H}_E)$. Again by linearity we extend the above definition for an arbitrary element from $\mathcal{L}^2(\mathcal{H}_S \ot \mathcal{H}_E)$. It should be stressed that $\mathcal{P}_\alpha$ define projectors in $\mathcal{L}^2(\mathcal{H}_S \ot \mathcal{H}_E)$ but not in $\mathcal{H}_S \ot \mathcal{H}_E$.
Now we apply Feshbach technique replacing $\mathcal{H}_S \ot \mathcal{H}_E$ by $\mathcal{L}^2(\mathcal{H}_S \ot \mathcal{H}_E)$: let $\zeta = \kappa_S \ot \kappa_E$ be an initial amplitude of the composed system. It is therefore clear that the initial state $\zeta\zeta^\dagger = \rho \ot \Omega$ is a product state, where $\rho = \kappa_S\kappa_S^\dagger$ and $\Omega = \kappa_E\kappa_E^\dagger$.
Let $\zeta_t \in \mathcal{L}^2(\mathcal{H}_S \ot \mathcal{H}_E)$ satisfy Schr\"odinger-like equation
\begin{equation}\label{}
    i \partial_t{\zeta}_t = V(t)\zeta_t\ .
\end{equation}
Performing  the same steps leading to formula (\ref{Lambda-S}) one arrives at
\begin{equation}\label{Lambda-G}
    \Lambda_t(\rho) = \mathcal{Z}_t \rho \mathcal{Z}_t^\dagger + {\rm Tr}_E ( \mathcal{Y}_t [\,\rho \ot \Omega] \mathcal{Y}_t^\dagger)\ ,
\end{equation}
where $\mathcal{Z}_t$ satisfies
\begin{equation}\label{RG}
     \partial_t \mathcal{Z}_t = -i {V}_{\rm eff}(t) \mathcal{Z}_t - \int_0^t \mathcal{M}_{t,s} \mathcal{Z}_s\, ds\ ,
\end{equation}
with the effective  time-dependent system Hamiltonian  defined by ${V}_{\rm eff}(t) = {\rm tr}_E(V(t) \, \mathbb{I}_S\ot \Omega)$, $\mathcal{M}_{t,s} = {\rm tr}_E(\mathcal{K}_{t,s} \, \mathbb{I}_S \ot \Omega)$ and $\mathcal{K}_{t,s} = \mathcal{V}_{01}(t) \mathcal{W}_{t,s} \mathcal{V}_{10}(s)$. Moreover, we introduced $\mathcal{V}_{ij}(t) = \mathcal{P}_i V(t) \mathcal{P}_j$.
The propagator $\mathcal{W}_{t,s}$ reads
\begin{equation}\label{}
    \mathcal{W}_{t,s} = \mathcal{T}\, \exp\left( -i \int_s^t \mathcal{V}_{11}(u) du \right)\ ,
\end{equation}
and finally the operator $\mathcal{Y}_t$ is defined by
\begin{equation}\label{}
    \mathcal{Y}_t = \int_0^t ds \ \mathcal{W}_{t,s} \mathcal{V}_{10}(s) (\mathcal{Z}_s \ot \mathbb{I}_E)  \ .
\end{equation}
Interestingly, the dynamical map defined in (\ref{Lambda-G}) has exactly the same form as in (\ref{Lambda-S}) with $Z_t$ replaced by $\mathcal{Z}_t$, $Y_t$ by $\mathcal{Y}_t$ and pure state $|\psi_E\>\<\psi_E|$ is replaced by an arbitrary mixed state $\Omega$ of the environment. Finally, the Hilbert space projectors $P_\alpha$ are replaced by the projectors $\mathcal{P}_\alpha$ acting in the space of amplitudes. It should be stressed that although  the dynamical map (\ref{Lambda-G}) is defined by the same formula  as (\ref{Lambda-S}) the derivation of $\Lambda_t$ with arbitrary mixed state of the environment was possible only after passing to the space of amplitudes.
Finally, let us observe that projectors $\mathcal{P}_\alpha$ acting in the space of amplitudes are gauge-dependent. However,
one proves that the corresponding dynamical map $\Lambda_t$ depends only upon the multi-time correlation functions of the environmental operators, and hence it is perfectly gauge-invariant.


{\em Conclusions. --} 
Contrary to the standard Nakajima-Zwanzig projection technique in the Banach space of density operators our general method is based on the Feshbach projection technique in the Hilbert space of amplitudes. The main advantages of presented approach are $i)$ it is based on a much simpler dynamical equation not for the dynamical map itself but for the linear operator acting in the system Hilbert space. Solving this equation one constructs the exact dynamical map.  $ii)$ This approach  enables one to work with mixed states of the environment. $iii)$ Its crucial property is the ability for coherent approximations leading to legitimate (completely positive and trace preserving) approximated dynamics. This is a big advantage with respect to the  Nakajima-Zwanzig approach. This is due to the fact that Feshbach technique works within the Hilbert space of vector states or amplitude operators. This is physically more intuitive and mathematically much simpler than the Nakajima-Zwanzig technique in the abstract Banach space of density operators.
We  illustrated our approach by the spin-boson model beyond RWA. It provides a key model in the theory of open quantum systems \cite{Breuer} due to the fact that it is exactly solvable (within RWA) if $\psi_E = |{\rm vac\>}$. Our approach shows that this feature corresponds to the fact that the standard spin-boson model is exact in the Born-like approximation. Adding counter-rotating terms and/or replacing $|{\rm vac}\>$ by another state (pure or mixed) make this model untractable within standard approach \cite{Breuer}. The generalized Feshbach projection technique enables one to deal  both with counter-rotating terms and arbitrary mixed state of the environment. The power of this method consists in the fact that one may deal with arbitrary mixed state of the environment by replacing the corresponding correlation functions.




{\em Acknowledgements --}  This work was partially supported by the National Science Centre project
DEC-2011/03/B/ST2/00136. Authors thank anonymous referees for many valuable comments.

\end{document}